\documentclass[manuscript]{aastex}

\usepackage{natbib}

\shorttitle{Phase Transition Generated Magnetic Field}
\shortauthors{Kahniashvili et al.}

\begin{document}

\title{Phase transition generated cosmological magnetic field at large scales}

\date{\today~~KSUPT-09/3}

\author{Tina Kahniashvili}
\email{tinatin@phys.ksu.edu} \affil{McWilliams Center for Cosmology
and Department of Physics, Carnegie Mellon University, 5000 Forbes
Ave, Pittsburgh, PA 15213} \affil{Department of Physics, Laurentian
University, Ramsey Lake Road, Sudbury, ON P3E 2C, Canada}
\affil{Abastumani Astrophysical Observatory, Ilia State University,
2A Kazbegi Ave., Tbilisi, 0160, Georgia}

\author{Alexander G. Tevzadze}
\email{aleko@tevza.org} \affil{Abastumani Astrophysical Observatory,
Ilia State University, 2A Kazbegi Ave., Tbilisi, 0160, Georgia}
\affil{Faculty of Exact and Natural Sciences, Tbilisi State
University, 1 Chavchavadze Ave., Tbilisi, 0128, Georgia}

\author{Bharat Ratra}
\email{ratra@phys.ksu.edu} \affil{Department of Physics, Kansas
State University, 116 Cardwell Hall, Manhattan, KS 66506}

\begin{abstract}

We constrain a primordial magnetic field (PMF) generated during a
phase transition (PT) using the big bang  nucleosynthesis bound on
the relativistic energy density. The amplitude of the PMF at large
scales is determined by the shape of the PMF spectrum outside its
maximal correlation length scale. Even if the amplitude of the PMF
at 1 Mpc is small, PT-generated PMFs can leave observable signatures
in the potentially detectable relic gravitational wave background if
a large enough fraction ($1-10\%$) of the thermal energy is
converted into the PMF.

\end{abstract}


\maketitle

\section{Introduction}

A cosmological seed PMF (generated during or prior to the
radiation-dominated epoch) has been proposed to explain the
existence of observed $\sim 10^{-6}-10^{-5}$ Gauss (G) magnetic
fields in galaxies and clusters \citep{Widrow,Vallee}. To preserve
approximate spatial isotropy a PMF has to be small and hence can be
treated as a first order term in perturbation theory. In the
standard cosmological model \citep{model} the energy density
parameter of a PMF, $\Omega_B = \rho_B/\rho_{\rm cr}$, is
significantly less than unity. Also, a PMF must be smaller than
those observed in galaxies ($10^{-5}$ G), so $\Omega_B h_0^2 <
10^{-4}$ where $h_0$ is the Hubble constant in units of 100 km
s$^{-1}$ Mpc$^{-1}$. Since the PMF energy density contributes to the
radiation field, the big bang nucleosynthesis (BBN) bound implies
$\Omega_B h_0^2 \leq 2.4\times 10^{-6}$. The ratio of $\rho_B$ and
the energy density of radiation $\rho_{\rm rad}$ is constant during
cosmological evolution, if the PMF is not damped by a MHD (or other)
process and so stays frozen into the plasma. Direct measurement of a
cosmological MF is based on the Faraday rotation effect. A potential
extension of this method, based on the rotation of the cosmic
microwave background (CMB) polarization plane, appears promising
\citep{kosowsky96,sethi03,camp04,kklr05,Giovannini08,kmk09}. In
addition, a PMF leaves imprints on the CMB temperature and
polarization anisotropies \citep[for a review see][]{cmb}. Recently
two different groups \citep{neronov,limit2} reported the detection
of a lower bound on the large-scale correlated MF amplitude, of
order $10^{-16}-10^{-15}$ G at 1 Mpc scale, through blazar
observations.

In this paper we consider cosmological PMFs generated by causal
processes during phase transitions (PTs) such as the electroweak
(EW) and QCD PTs \citep[][and references
therein]{pmf1,pmf2,pmf3,pmf4,pmf5,pmf6,pmf7,pmf8,pmf9,pmf10,pmf11}.
The main parameters of interest are the temperature $T_\star$ and
the number of relativistic degrees of freedom $g_\star$ when the PMF
is generated. We only use fundamental physical laws, such as
conservation of energy, and how the magnetic field interacts with
the cosmological plasma through MHD turbulence, and do not make any
assumption about the physical processes responsible for PMF
generation. In contrast to earlier studies that estimate current
limits on the PMF, we employ only energy conservation arguments
without referring to specific processes responsible for transforming
energy from a PMF to gravitational waves or thermal or other forms
of energy.  We discuss cosmological signatures of such a PMF,
including effects on the CMB temperature and polarization
anisotropies and the production of gravitational waves (GWs). We
employ natural units with $\hbar = 1 = c$ and gaussian units for
electromagnetic quantities.

In Sec.\ 2 we define the spatial and temporal characteristics of the
PMF and derive the BBN limits expressed in terms of the PT
parameters. Results of our analysis are presented in Sec.\ 3, where
we discuss the effects of a magnetic field on the CMB temperature
and polarization anisotropies and relic gravitational waves.
Conclusions are presented in Sec.\ 4.

\section{Magnetic Field Spectrum}

The maximal correlation length $l_{\rm max}$ for a causally
generated PMF cannot exceed the Hubble radius at the time of
generation, $H_\star^{-1}$. Hence $\gamma= l_{\rm max}/ H_\star^{-1}
\leq 1$, where $\gamma$ can be associated with the number of PMF
bubbles within the Hubble radius, $N \propto \gamma^3$. The comoving
length (measured today) corresponding to the Hubble radius at
generation is inversely proportional to the temperature $T_\star$,
\begin{equation}
\lambda_{H_\star} = 5.8 \times 10^{-10}~{\rm
Mpc}\left(\frac{100\,{\rm GeV}}{T_\star}\right)
\left(\frac{100}{g_\star}\right)^{{1}/{6}}, \label{lambda-max}
\end{equation}
and is equal to 0.5 pc for the QCDPT (with $g_\star=15$ and
$T_\star=0.15$ Gev) and $6 \times 10^{-4}$ pc for the EWPT (with
$g_\star=100$ and $T_\star =100$ GeV), and the comoving PMF
correlation length $\xi_{\rm max} \leq \lambda_H$. This inequality
only accounts for the expansion of the Universe while ignoring any
possible effects of MHD turbulence (free turbulence decay or an
inverse cascade if a helical PMF is present); this issue is
addressed below.

If generated prior to BBN, the maximal value of the PMF energy
density must satisfy the BBN bound, i.e., the total energy density
of the PMF at nucleosynthesis $\rho_B(a_{\rm N})$ (where $a_{\rm N}$
is the scale factor at nucleosynthesis) should not exceed 10$\%$ of
the radiation energy density $\rho_{\rm rad}(a_{\rm N})$. We first
consider the simplest case when the PMF energy density evolves just
due to the expansion of the Universe. The conservation of the
magnetic flux in this case leads to $\rho_B (t) \propto 1/a^4(t)$.
Thus the ratio $\rho_B/\rho_{\rm rad}$ is constant, and the BBN
constraint can be expressed as $\rho_B/\rho_{\rm rad}(a_\star) \leq
0.1 $. Simple algebra gives that the maximal comoving value of the
effective PMF $B^{(\rm eff)}=\sqrt{8\pi \rho_B}= 8.4 \times 10^{-7}
(100/g_\star)^{1/6}$ G, if no PMF damping occurs before BBN. Even if
the PMF energy is converted to another field contributing to the
radiation \citep[for example, GWs,][] {g1,gw}, there is only
$\rho_B(a_\star)$ magnetic energy available. The next issue is to
determine how this energy is distributed at different wavelengths,
and the comoving PMF at a given comoving length scale $\lambda$. In
this paper we address this issue based on a phenomenological
description of MHD processes in the early Universe.  Of course, for
a scale-invariant \citep{ratra92,bamba08} or homogeneous PMF the
limit remains the same at any scale. For any other PMF spectrum the
limits are strongly scale dependent. As we will show below, the
total PMF energy density is the physical quantity that has
cosmological relevance. Obviously, from a limit on the PMF energy
density we are able to reconstruct the effective value of the
magnetic field amplitude. Note that the maximal value of the PMF
(from the BBN bound) is independent of the temperature at generation
$T_*$, and depends only very weakly on the number of relativistic
degrees of freedom at the moment of generation.

In this paper we propose treating the initial PMF energy density
$\rho_B$ as the magnetic energy density injected into the
cosmological plasma at the comoving length scale $\lambda_{0}$. We
justify our assumption that $\lambda_0$ should be identified as the
size of the largest magnetic eddy by noting that the PMF is involved
in MHD processes driven by turbulence. The typical length scale of
the turbulence is determined by the PT bubble size. After
generation, PMF evolution (during the PT) depends sensitively on the
length scale under consideration. If the relevant time scales are
shorter than $H_\star^{-1}$ we can neglect the expansion of the
Universe.\footnote{In the ideal case there is no magnetic energy
damping due to the redistribution of the initial magnetic energy
density $\rho_B(t_\star) $ through the different wavelength. To
distinguish the initial magnetic energy present at a given scale
from the total magnetic energy density right after redistribution,
we define the magnetic energy density ${\mathcal E}_M $ and it is
obvious that ${\mathcal E}_M \leq \rho_B$. } In this case, we must
distinguish three sub-Hubble-radius regimes: $k_{H_\star} < k < k_0$
(where $k_{H_\star}=2\pi/\lambda_{H_\star}$ and
$k_0=2\pi/\lambda_0$; the large scale decay regime); $k_0<k<k_d$,
with $k_d=2\pi/\lambda_d$ the damping wavenumber scale related to
plasma properties (the turbulence regime); and, $k>k_d$ (the viscous
damping regime). The interaction of the PMF with the plasma, and as
a consequence the dynamics of the PMF, is sensitive to the presence
of magnetic helicity \citep[see][for magnetic helicity generation
mechanisms] {hel1,hel2,hel3,hel4,pmf9,pmf10}. The expansion of the
Universe leads to additional effects. Most important among these are
the additional damping of the PMF and the different time dependence
of the growth rates for the comoving length scales ($L \propto a$)
and the Hubble radius scale ($H^{-1} \propto t$, where $t$ is
physical time). As a result, if a perturbation with wavenumbers $k$,
$k_{H_0}< k < k_{H_\star}$, was outside the Hubble radius, at some
time this perturbation will enter the Hubble radius \citep{BN}.

The magnetic energy $E_M(k,t)$ and helicity $H_M(k,t)$ density power
spectra are related to the magnetic energy and helicity densities
through $ {\mathcal E}_M(t) = \int^\infty_0 \! dk E_M(k,t) $ and $
{\mathcal H}_M(t) = \int^\infty_0 \! dk H_M(k,t)$. The magnetic
correlation length $ \xi_M(t) = [{\int^\infty_0 \!dk k^{-1}
E_M(k,t)}]/{{\mathcal E}_M(t)}$ corresponds to the largest eddy
length scale. All configurations of the MF must satisfy the
``realizability condition'' \citep{B03,B03-1}: $|{\mathcal H}_M(t)|
\leq 2 \xi_M(t) {\mathcal E}_M(t)$. Also, the velocity energy
density spectrum $E_K(k,t)$ is related to the kinetic energy of
turbulent motions through $ {\mathcal E}_K(t) = \int^\infty_0 \! dk
E_K(k,t).$

To account for the expansion of the Universe we make use of the fact
that conformal invariance allows for a description of MHD processes
in the early Universe by simply rescaling all physical quantities in
terms of their comoving values and using the conformal time $\eta$
\citep{pmf3,jedamzik1,jedamzik2}. After this procedure the MHD
equations include the effects of the expansion while retaining their
conventional flat spacetime form. To keep the description as simple
as possible we work with dimensionless quantities, such as the
already defined parameter $\gamma$ and the normalized wavenumber and
normalized energy density defined below.

\subsection{Magnetic field spatial structure}

We first consider the non-helical case. For large enough Reynolds
number the magnetic energy is re-distributed by a Kolmogorov
turbulence direct cascade.  From the analogy between the Kolmogorov
laws for hydrodynamic and magnetic turbulence, the magnetic energy
dissipation comoving rate per unit enthalpy is $\varepsilon_M \simeq
(2/3)^{3/2} k_0 v_A^3$, with $v_A = \sqrt{1.5 \rho_B/\rho_{\rm
rad}}$ being the effective Alfv\'en velocity corresponding to the
total fluid-injected PMF energy, i.e.\ $\varepsilon_M = k_0
(\rho_B/\rho_{\rm rad})^{3/2}$. In the absence of PMF damping at
$k=k_0$, for Kolmogorov turbulence, $E_M(k)=C_M \rho_B k_0^{-1}
{\bar k}^{-5/3}$ when $k_0 < k < k_d$, where $C_M$ is a constant of
order unity ($C_M = 1.3 C_K $, where $C_K \simeq 1$ is the
Kolmogorov constant),  and ${\bar k} =k/k_0$ is the normalized
wavenumber. At large scales when $k<k_0$ we model the PMF energy
spectrum by a power law, $E_M(k) \propto k^\alpha$. Requiring
continuity of the PMF spectrum at $k=k_0$, $E_M(k) = C_M \rho_B
k_0^{-1} {\bar k}^\alpha$ for $k<k_0$. It is natural to assume that
the MF energy injection scale $\lambda_0$ is the same as the maximal
correlation length of the PMF, i.e.\ $\lambda_0 \simeq l_{\rm max}
a_0/a_\star$.

The spectral index $\alpha$ of the largest scale MF energy density
has been much discussed. \citet{hogan} requires causality of the
field and argues that the PMF energy density spectrum must be white
noise for scales larger than the causal horizon, $\lambda_0$. This
corresponds to $\alpha=2$. \citet{caprini-1} claim that this
violates the divergence-free PMF requirement and instead demand
$\alpha=4$. Both of these spectra, $\alpha=2$ (Saffman) and
$\alpha=4$ (Batchelor),{\footnote{$E \propto k^4$ is sometimes
called the von K\'arm\'an spectrum.} are well known in the
turbulence literature and, as discussed in \citet{textbooks}, their
realization depends on initial conditions.
In \citet{kbtr10} the magnetic field spectral shape at large scales
was obtained from direct 3D numerical simulations and $\alpha \in
(2, 4)$ was found, depending on the initial conditions.
Another possibility is Kazantsev's $\alpha = 3/2$ value, which can
be rapidly achieved during the turbulence decay process discussed in
\citet{Axel-1} and \citet{Axel-12}. To keep the analysis as general
as possible, we keep $\alpha$ arbitrary as much as possible.

Requiring ${\mathcal E}_M \leq \rho_B$ we obtain $C_M \leq 2 (\alpha
+1)/(3\alpha+5)$. With the nucleosynthesis requirement that the
injected magnetic field energy density must be smaller than the
10$\%$ of the relativistic energy density, $\rho_B \leq 0.1
\rho_{\rm rad}$,\footnote{We can extend the 10$\%$ bound to epochs
earlier than nucleosyntheses. This is justified if the PMF was not
damped away or MF energy was not transformed to another relativistic
component, e.g.\ gravitational radiation (Caprini \& Durrer 2001).}
and neglecting the contribution to the energy density from scales
smaller than the viscous damping scale $\lambda_d$, for the maximal
allowed value of $C_M$, we have $ C_M \rho_B \leq 2.81 \times
10^{-14}C_{M, {\rm max}} (100/g_\star)^{1/3}$ G$^2$, and the
magnetic energy spectrum
\begin{equation}
E_M({k} )  \leq   \frac{5.2
(\alpha+1)}{3\alpha+5}~\left(\frac{100\ {\rm GeV}}{T_\star}\right)
\left(\frac{100}{g_\star}\right)^{1/2} \gamma ~
\frac{(10^{-9}~{\rm G})^2} {{\rm pc }^{-1}} \left \{
\begin{array}{c}
{\bar k}^{\alpha} ~~~~~ {\rm if} ~~~ {\bar k}<1 \\
{\bar k}^{-5/3} ~~ {\rm if} ~~~ {\bar k} > 1
\end{array}
\right. . \label{em}
\end{equation}
Note that plasma properties determine the damping wavenumber $k_d$
that is defined by the Reynolds number ${\rm Re} \gg 1$ of the fluid
during the PT, $k_d = k_0 {\rm Re}^{3/4}$. The plasma viscosity and
other characteristics change during the evolution of the Universe,
and also are scale dependent. I.e., the Reynolds number is time and
scale dependent, as shown in Fig.\ 7 of Caprini et al.\ (2009b).
Equation (\ref{em}) accounts for the expansion of the Universe. If
the perturbation wavenumber is inside the inertial regime
$k_0<k<k_d$, it will always stay inside this regime, independent of
the expansion.

Defining $B_\lambda$ as a smoothed PMF over a sphere of radius
$\lambda $ ($\lambda >\lambda_0$) we have for the MF energy density
on scales larger than the maximal correlation length, ${\mathcal
E}_M^{~\rm LS}$ $=\int_0^{k_0} dk E_M({\bar k})$ $= {B_\lambda^2
(k_0\lambda)^{\alpha+1}}/[8\pi \Gamma(0.5\alpha +1.5)]$, where
$\Gamma$ is the Euler Gamma function \citep{kr07}. This leads to the
upper bound on $B_\lambda$,
\begin{eqnarray}
\frac{B_\lambda}{10^{-9}{\rm G}} \left(\frac{\lambda}{1~{\rm
Mpc}}\right)^{(\alpha+1)/2} & \leq & {2.28 \times 10^{-5\alpha -3}
\over \sqrt{3\alpha+5}} \gamma^{(\alpha+1)/2}  \left[8\pi
\Gamma\left(\frac{\alpha+3}{2} \right)\right]^{1/2} \nonumber\\ &&
\left(\frac{100~{\rm GeV}}{T_\star}\right)^{(\alpha + 1)/2}
\left(\frac{100}{g_\star}\right)^{(\alpha+3)/12},  \label{bl}
\end{eqnarray}
shown in Fig.\ 1, for the EWPT with $T_\star =100$ GeV and
$g_\star=100$ (upper panel) and the QCD PT with $T_\star = 150$
MeV and $g_\star = 15$ (lower panel). The parameter $\gamma$ is
related to the phase transition duration parameter $\beta$ and
the phase transition bubble wall speed $v_b$ as $\gamma = v_b
\beta^{-1} H_\star$, and for a strong enough phase transition,
with $v_b \rightarrow 1$, $\gamma \simeq \beta^{-1} H_\star$.
Equation (\ref{bl}) is based on energy conservation and accounts
only for the expansion of the Universe. While previous studies
\citep{hogan,jedamzik1,jedamzik2,caprini09b,cds09} address
magnetic field limits, the result shown here is independent of
the magnetic field generation model details, as well as those
that govern the following magnetic energy transformation process
dynamics. For the first time, we directly connect the magnetic
field strength limit to the characteristics of the PT: the bubble
wall velocity $v_b$ and the duration of the PT. Both these enter
through the parameter $\gamma$ which is dimensionless and
independent of the expansion. Note that Eq.\ (\ref{bl}) and, as a
consequence, Fig.\ 1 assume that $\rho_B (a_\star) \leq 0.1
\rho_{\rm rad} (a_\star)$, thus any MF energy dissipation is
neglected. In this sense, nucleosynthesis bound will be lower
compared to the values shown in Fig.\ 1, since any realistic
calculation should include dissipation effects that can
significantly reduce the MF amplitudes. On the other hand, Fig.\
1 is valid if the MF energy density is converted into a
gravitational wave signal; also see
\citet{caprini09b,cds09}.\footnote{Note that a limit qualitatively
similar to that above can be obtained from considering the
backreaction of the MF. A strong enough PMF will violate the
isotropy of the Friedmann-Lema\^itre-Robertson-Walker metric. To
be able to consider the PMF energy density as a first order
perturbation with respect to the radiation energy density at the
moment of PMF generation, it is not unreasonable to assume that
$\rho_B(t_\star) \leq 0.1 \rho_{\rm rad}$. We thank the referee
for emphasizing this point.}

For the inflation generated PMF ($\alpha \rightarrow -1$) the
dependence on $T_\star$ and $g_\star$ disappear, and $B_\lambda \leq
10^{-6}$ G.  As emphasized above the PMF limits shown in Fig.\ 1 do
not account for the non-cosmological evolution and damping of the
PMF;\footnote{ Note that the simple dilution of the PMF due to the
expansion of the Universe is accounted for.} they address only the
spatial structure of the PMF at large scales constrained by
$\rho_B(t_\star) \leq 0.1 \rho_{\rm rad}$. On the other hand, PMF
damping would reduce the amplitude of the magnetic field, so Fig.\ 1
gives a real upper limit on the PMF. To avoid confusion we emphasize
that these limits cannot be improved by the constraint from
overproduction of GWs \citep{gw}, since only $\rho_B$ is available
(at the moment of the PMF generation) to be transformed into another
form of energy. The observable signatures of a PMF are also
determined by the PMF total energy density (see Sec.\ III) and in
this case the non-cosmological evolution of the PMF must be taken
into account.  Based on dimensional analysis we address the time
evolution of the PMF in the next subsection, with more precise
numerical results presented in \citet{kbtr10}.

\subsection{Magnetic field temporal characteristics}

The PMF spectrum is characterized not only by its spatial
distribution, but also by its characteristic times: i) the
largest-size eddy turn-over time $\tau_0 \simeq l_0/v_A$, which can
also be used to determine the minimal duration of the source needed
to justify use of the stationary turbulence approximation
\citep{P52,P52-2}; ii) the turbulence direct cascade time-scale
$\tau_{\rm dc}$; and, iii) the large-scale turbulence decay time
$\tau_{\rm ls}$.

We first consider the inertial range for which two timescales,
$\tau_0$  and $\tau_{\rm dc}$, are relevant. For the EWPT ($\gamma
\simeq 0.01$), and for reasonable moderate values of the Alfv\'en
velocity ($v_A \leq 0.3$), the maximal size magnetic eddy turn-over
physical timescale $\tau_0 \simeq \gamma H_\star^{-1}/v_A $ is
significantly shorter than the Hubble time $H_\star^{-1}$. As the
Universe expands, the eddy turn-over time grows as $\tau_{\rm tr}
\propto a$, since $v_A$ is a time-independent parameter and all
length scales increase. The increased turn-over timescale makes all
MHD processes slower. We can effectively describe the expansion
effects by considering MHD turbulence with a typical velocity that
is decreasing as $\propto a^{-1}$. The time characteristic
$\tau_{dc}$ is determined by the proper dissipation rate ${\bar
\varepsilon}_M $,\footnote{Since, for the developed turbulence case,
the magnetic energy proper dissipation rate per unit enthalpy must
be approximately equal to the mean energy input rate per unit
enthalpy of the source, i.e., the turbulence cascade time scale
$\tau_{\rm dc} \simeq 2\pi {\bar \varepsilon_M}^{-1} (\rho_B/{\rm
w}_{\rm rad}) \simeq \tau_0$, where w$_{\rm rad}$ is the radiation
enthalpy \citep{kmk02}.} and it also increases as the Universe
expands. MHD turbulence decorrelation is a complex process
\citep{ts07} that is currently not fully understood. To proceed we
employ Kraichnan's approach \citep{K64} and specify decorrelation
function $f_{\rm dc}[\kappa(k_{\rm ph}),\tau] = \exp \! \left[-\pi
\kappa^2(k_{\rm ph}) \tau^2/4 \right]$ defined within the inertial
range, $k_0<k<k_d$. Here $\tau$ is the duration of the turbulence
process, and $\kappa(k_{\rm ph}) = {{\bar
\varepsilon}_M^{1/3}}k_{\rm ph}^{2/3}/{\sqrt{2\pi}}$ where $k_{\rm
ph}$ is the physical wavenumber related to the comoving $k$ through
$k_{\rm ph}(a) = ka_0/a$ ($a_0$ is the value of the scale factor
now). Hence, we have,
\begin{equation}
f_{\rm dc}[{\bar k},\tau] = \exp \! \left[-\frac{ 2 \pi^2}{9} \left(
\frac{\tau}{\tau_0} \right)^2 {\bar k}^{4/3} \right] ~. \label{dec}
\end{equation}
For non-helical turbulence, by considering the largest-size magnetic
eddy decorrelation, $\tau_{dc} \simeq 0.5 \tau_0$, thus the direct
cascade time scale is much shorter than the Hubble time, and this
justifies the neglect of the expansion of the Universe, as well as
the assumption made above to neglect the energy density for
$\lambda<\lambda_0$. Figure \ref{Em_decorr} shows the normalized PMF
spectral energy density decays within one turnover time for
perturbations within the inertial range.
Note that the results presented in Fig.\ 2 can not be directly
compared with those in \citet{kbtr10}, since Eq. (\ref{dec}) assumes
high enough Reynolds numbers while the numerical simulations are
conducted at finite resolution leading to unrealistic, low Reynolds
numbers.

Another characteristic time is related to the decay of large-scale
turbulence. Large scale turbulence processes are much slower than
those occurring at smaller scales within the inertial range.
Specific to this process is that there is no magnetic or
hydrodynamic turbulence production source and free decay occurs. In
this case, for consistency, we continue to use conformal time
$\eta=\int dt/a$ and the normalized and comoving quantities.
In the case of laboratory turbulence, to study large-scale decay one
frequently assumes the grid turbulence decay law ${ k}(t) \propto
t^{-1.3}$ for perturbations with wavenumber $k (t) < k_G$, where
$k_G=2\pi/\lambda_G$ is the wavenumber corresponding to the grid
scale $\lambda_G$ that can be associated with the correlation length
\citep[see, e.g.][]{pope}. In the expanding Universe the correlation
length $\xi_M (\eta) $ and the Hubble radius length $\lambda_H
(\eta) \simeq H^{-1}$ set natural length scales which are the
analogue of the grid size in laboratory turbulence. On the other
hand the decay law in the expanding Universe, according to
\citet{kbtr10}, is substantially slower than in the laboratory case,
i.e. $k(\eta ) \propto \eta^{-1/2}$. A similar difference holds for
the time evolution of the PMF energy density: the numerical
simulations show $E_M({\bar k}, \eta) \propto \eta^{-1}$ while the
grid turbulence approach gives $E_M({\bar k}, t) \propto t^{-n_G}$
with $n_G \in (1.13, 1.25)$ \citep{frisch75,biskamp93}.

Figure \ref{Em_kt} shows the spatial and temporal surface of freely
decaying turbulence at large scales ignoring MHD dynamo effects and
neglecting all processes within the inertial range, ${\bar k} >1$.
Free decay is important only at large scales. As well as the grid
turbulence analogy, the direct numerical simulations shown  that the
PMF spectral energy density decreases much slower than the
turbulence driven by the direct cascade. After free decay reduces
the MF power substantially, the PMF on scales $k \ll k_0$ can be
treated as a MF that is unaffected by turbulence. Of course, when
considering realistic cosmological turbulence, in contrast to the
laboratory turbulence case, the evolution of the fluid viscosity
must be taken into account, which can change the PMF correlation
length and energy density scaling laws shown here; see
\citet{caprini09b} for a different model of free-decaying MHD
turbulence. We also argue \citep{kbtr10} that free decay laws are
very sensitive to the initial conditions, i.e.,\ the free decay law
depends on how the PMF was generated \citep[][and references
therein]{pmf1,pmf2,pmf3,pmf4,pmf5,pmf6,pmf7,pmf8,pmf9,pmf10,pmf11};
whether the PMF was generated through bubble collisions, leading to
$l_0 \simeq v_b \beta^{-1} $ with $v_b$ the bubble wall velocity and
$\beta $ the bubble nucleation rate parameter ($\beta \simeq 100
H_\star$ for the EWPT) \citep{kks09}, or if the PMF was present
prior to the PT \citep{ratra92,bamba08}. Another uncertainty comes
from the kinetic (vortical) energy density spectrum $E_K$
\citep{Axel-1,Axel-12}.

The presence of even a small amount of magnetic helicity
substantially affects PMF evolution
\citep{BN99,BN991,BN992,axel,jedamzik1,jedamzik2,campanelli}. If
there is only a little magnetic helicity, first a direct cascade
develops. At the end of this first stage the turbulence relaxes to a
maximally helical state \citep{axel,jedamzik1} that satisfies
$|{\mathcal H}_M(\eta)| \leq 2 \xi_M(\eta) {\mathcal E}_M(\eta)$ and
the second inverse-cascade stage starts. Conservation of magnetic
helicity implies that the magnetic energy density decays in inverse
proportion to the correlation length growth during the inverse
cascade. In contrast, for the case of well-established non-helical
turbulence, the effect of magnetic helicity is still under
discussion \citep{BN99,BN991,BN992,axel,jedamzik1,campanelli}. The
main point of debate is related to the magnetic correlation length
growth rate during the inverse cascade, i.e.\ $\xi_M(\eta) \propto
\eta^{n_\xi}$, where the index $n_\xi$ is argued to be $1/2$
\citep{BN99,BN991,BN992,axel} or $2/3$ \citep{jedamzik1,campanelli}.
The total magnetic energy density ${\mathcal E}_M(\eta) \propto
\eta^{-n_\xi}$, and the decay of large-scale kinetic energy
${\mathcal E}_K(\eta)$ (and as a consequence the ratio between the
magnetic and kinetic energy densities ${\mathcal E}_M(t)/{\mathcal
E}_K(\eta)$) are sensitive to $n_\xi$. In particular,
\citet{BN99,BN991}, \citet{BN992}, and \citet{axel} argue that
${\mathcal E}_K(\eta) \propto \eta^{-1}$, implying a faster decay of
kinetic energy at large scales, while the results of
\citet{jedamzik1} and \citet{campanelli} lead to a constant
${\mathcal E}_M(\eta)/{\mathcal E}_K(\eta)$ within the inverse
cascade and ${\mathcal E}_K(\eta) \propto \eta^{-2/3}$.

To compute the decay rate in physical time we use $t \propto
\eta^2$. As in the case of non-helical turbulence, free decay here
occurs over a longer period than when ignoring the expansion of the
Universe. Even so, in both these models, with $n_\xi=1/2$ and $n_\xi
= 2/3$, the kinetic turbulent energy density significantly decays on
the EWPT timescale $H_\star^{-1}$, freezing the MF in the plasma,
see Fig.\ 4.

\section{Results and Discussion}

The limits on the PMF at large scales are much stronger for the
non-helical turbulence case; without the decay the constraint at
zero redshift on 1 Mpc is $10^{-28}$ G for $\alpha=4$ for an EWPT
generated PMF, see Fig.\ 1 \citep[also see][]{caprini09b}. In the
$\alpha=3/2$ case, the PMF can reach values of order
$10^{-12}-10^{-11}$ G that are required for seed MFs that might be
able to explain observed MFs in galaxies and clusters \citep{Dolag}.
Again, the limits above correspond to the case when the PMF is
frozen into plasma and no MHD processes are accounted for, Of
course, large-scale decay of turbulence will strengthen these limits
for both the non-helical and helical cases. On the other hand,
accounting for large-scale PMF decay the BBN bound does not imply
$\rho_B \leq 0.1 \rho_{\rm rad}$ when the PMF is generated. However,
there is another requirement: the PMF cannot be the only component
during the radiation-dominated epoch, thus $\rho_B/\rho_{\rm
rad}(T_\star) <1 $. Even though our analysis is preliminary, it
seems that a PT generated PMF requires an effective amplification
mechanism (such as a dynamo), or a specific initial condition, to
act as a viable seed field for observed MFs in galaxies and
clusters. Non-linear processes during the PT can be responsible for
the change of shape of the PMF at large scales and substantially
long-duration turbulent sources can lead to significant magnetic
power at large scales \citep{kbtr10}.

A PT generated PMF may have observable cosmological signatures. In
what follows we examine two of them, CMB anisotropies and relic
gravitational waves. Before addressing the PMF cosmological
signatures, let's consider the effective magnetic field  and the
magnetic field damping scale.

\subsection{The effective magnetic field and the Alfv\'en damping scale}

As noted in Sec.\ 1, a stochastic MF can be described also by the
effective magnetic field value, $B^{(\rm eff )}$. Simple
calculations allows us to connect both quantities, the smoothed
magnetic field $B_\lambda$ and $B^{(\rm eff )}$, through the
following relation,
\begin{equation}
B^{(\rm eff )} = \frac{B_\lambda (k_D
\lambda)^{(\alpha+1)/2}}{\sqrt{\Gamma(\alpha/2+3/2)}}. \label{b-eff}
\end{equation}
where $k_D$ is the wavenumber cut-off above which the magnetic
energy density spectrum $E_M(k)$ vanishes. In Sec.\ 2 we assumed
that the PMF spectrum vanishes due to plasma viscosity, and so we
considered a damping wavenumber $k_d$ determined by the Reynolds
number.

On the other hand, there are different MHD processes that result in
MF damping. In particular, \citet{cmb-1,cmb-12} and \citet{olinto}
study the damping of a homogeneous MF assuming the main dissipation
process is Alfv\'en wave viscosity, resulting in $k_D^{-1} \simeq
L_S v_A$, with $L_S$ being the Silk damping scale at recombination
and $v_A$ determined by the background homogeneous PMF. For the case
of a stochastic PMF \citep{cmb-21},\footnote{\citet{cmb-21} defines
the effective MF as the PM ${\bar B}_0$ smoothed over the damping
length scale $L_D= 2\pi/k_D$, so
\begin{equation}
{\bar B}_0 = B_\lambda \left( \frac{k_\lambda}{k_D}
\right)^{\frac{\alpha+1}{2}}
\end{equation}
while in the present paper $B^{({\rm eff})}$ is determined through
the total energy of the MF. The difference is of order of few due to
the prefactor $\sqrt{\Gamma(\alpha/2 +3/2)}$ with $\alpha \in (-3,
2)$; see below.} define $k_D$ as
\begin{equation}
\left({k_D \over {1 {\rm Mpc}}^{-1}}\right)^{\alpha + 3} \approx
2\times 10^4 \left({B_\lambda\over 10^{-9}\,{\rm G}}\right)^{-2} h
\left({k_\lambda\over {\rm Mpc}^{-1}}\right)^{\alpha + 1},
\label{damping-scale}
\end{equation}
so $k_D$ becomes $B_\lambda$, $\lambda$, and $\alpha$ dependent.
Such a description must be used with caution: The picture of
Alfv\'en wave induced dissipation requires that the effective
background field be larger than the Alfv\'en wave associated field.
On the other hand, when $\alpha \geq 0$, $B_\lambda $ with $\lambda
> \lambda_D$ is significantly smaller than that associated with the
fluctuating Alfv\'en wave MF, so we can not treat the MF smoothed
over the large scale $\lambda$ as the effective background field
even when the length scale $\lambda$ is comparable with the current
Hubble radius. To avoid any possible confusion, it is useful rewrite
Eq. (\ref{damping-scale}) in a form independent of the $\lambda$
length scale and the smoothed value of the MF, $B_\lambda$. Using
Eq.~(\ref{b-eff}), it is easy to see that
\begin{equation}
\frac{k_D}{1{\rm Mpc}^{-1}} = 1.4 \sqrt{\frac{(2\pi)^{\alpha+1}
h}{\Gamma\left(\frac{\alpha+3}{2}\right)}} \left(
\frac{10^{-7}{\rm G }}{B^{({\rm eff})}} \right). \label{rho1}
\end{equation}
In the case of the scale-invariant spectrum any dependence on the
spectral index disappears (Eq.~(\ref{b-eff})), and $B^{({\rm eff})}
= B_\lambda$. For the cut-off scale we have the following simple
expression,
\begin{equation}
k_D = 1.4~{\rm Mpc}^{-1} h^{1/2} \left( \frac{10^{-7} ~{\rm
G}}{B^{({\rm eff})}} \right).
\end{equation}
The BBN limit ($B_{\rm eff} \leq 8.4 \times 10^{-7}$ G)  can be used
to put an upper limit on the cut-off wavenumber scale in the case of
the scale-invariant spectrum (i.e.\ with $n \rightarrow -3$), $k_D
\geq 0.2 $ Mpc$^{-1}$ ($\lambda_D  \leq 36$ Mpc). For an extremely
strong field that satisfies the BBN limit and with a white noise
spectrum, $\alpha=2$ \citep{hogan}, we have that $\lambda_D $ can be
large as 2.7 Mpc. Of course in this case it is inappropriate and
unjustified to consider a PMF at 1 Mpc as is conventional, since at
1 Mpc there is no MF.

An important question is how physical is a smoothed MF? At first
glance it seems that it reflects the strength of the MF at a given
scale. On the other hand, it is a tough task to compare $B_\lambda$
with observations. What the observations give is a measure of the MF
in terms of effects it induces and the correlation length of MF. It
is obvious that the smoothing scale and the correlation length are
two different quantities. MF induced effects are not directly
related to $B_\lambda$, rather they are strongly dependent on the MF
spectral shape, correlation length, etc.

Figure 1 shows the smoothed MF limits. One attempt to rule out a PMF
with spectral shape $\alpha = 4$ was based on the extremely strong
constraints on $B_\lambda$ at $\lambda=1$ Mpc \citep{gw}. In some
sense $B_\lambda $ reflects the normalization of the PMF and it is
strongly model dependent (it depends on the choice of $n_B$ and
$\lambda$), while the effective MF value $B^{(\rm eff )}$ is a
physical quantity that determines not only the total energy of MF
but also the cosmological signatures of a PMF.

\subsection{CMB temperature and polarization anisotropies}

A PMF induces CMB anisotropies. Usually, when considering PMF limits
imposed by CMB data, one refers to the amplitude of the smoothed PMF
$B_\lambda$ on large scales, typically $\lambda = 1$ Mpc
\citep{cmb}. On the other hand, when computing PMF induced CMB
temperature and polarization anisotropy power spectra one finds $C_l
\propto (B_\lambda^2 \lambda^{\alpha+1})^2 $
\citep{kr07,cmb-1,cmb-12,cmb-21}, and when considering Faraday
rotation of the CMB polarization plane the rotation angle and the
resulting B-polarization power spectra are $\propto  B_\lambda^2
\lambda^{\alpha+1}$ \citep{kklr05,kmk09}. From the definition of
$B_\lambda$ above, it is clear that PMF imprints on CMB fluctuations
are determined by $\Omega_B$ (i.e. $[B^{({\rm eff})}]^2$) or
$\Omega_B^2$ (i.e. $[B^{({\rm eff})}]^4$  ) \citep[also see][]
{olinto,sb}.

On the other hand, the MF energy density varies due to the
cosmological expansion, resulting in the time dependence of the
damping wavenumber $k_D$. During a PT that generates the MF, or when
the MF starts to interact with the plasma (if it is generated prior
to the PT), the total MF energy density at large scales is
determined by the wavenumber of the peak of the MF spectral energy
density, $k_0$. The value of this peak is fixed by the maximal
correlation length of the MF at the phase transition,
$\xi_M(\eta_star) = 2\pi/k_0$. In the simplest case that accounts
only for the expansion of the Universe and neglects all possible
effects of magnetic helicity (the inverse cascade) or/and free
turbulence decay, $\xi_M$  is independent of the evolution of the
Universe. In this case during a PT all wavenumber $k<k_0$ modes
contribute to the total MF energy density. The real time-evolution
is different: (i) in the absence of magnetic helicity the
correlation length increases (i.e.\ the corresponding peak
wavenumber $k_{\rm peak} (\eta) = 2\pi/\xi_M(\eta)$ shifts toward
smaller values) due to large-scale turbulence decay $\xi_M (\eta)
\propto \eta^{1/2}$, see Sec.\ 2.2; (ii) the presence of magnetic
helicity might make the correlation length grow faster, resulting in
$\xi_M (\eta) \propto \eta^{2/3}$, see Sec.\ 2.2. While the
correlation length increases, the energy density of the MF
decreases. Additional damping is caused by Alfv\'en wave
dissipation. The combined effect is that near the last-scattering
surface some MF modes have been damped and dissipated and so the
total MF energy density accounts for all modes with wavenumbers
$k<k_D$.

The Alfv\'en wave damping scale $k_D \ll k_0$ and thus only a small
part of the initial MF energy contributes to the CMB anisotropies.
In particular, the CMB anisotropies are determined by $\rho_{\rm
CMB} = (B^{\rm eff})^2 /(8\pi) \propto B_\lambda^2 (k_D
\lambda)^{\alpha +1}$. The CMB anisotropies of the scale-invariant
PMF (with $\alpha \simeq -1$) are determined only by the amplitude
of the PMF and are independent of $\lambda$ and $k_D$. A
scale-invariant PMF with amplitude larger than $10^{-9}$ G might
leave observable CMB anisotropy traces \citep{Yamazaki:2010nf}. The
situation is significantly different for a PMF with  $\alpha > -1$.
In this case the main contribution to the MF energy density comes
from small scales, and naively it might be expected that a PMF field
with a smoothed amplitude at 1 Mpc lower than $10^{-9}$ G may not
leave observable traces. In reality the situation is more
complicated. Even when the amplitude of the PMF is small enough, the
contribution to the CMB anisotropies is not linearly dependent on
$B_\lambda$. In  particular, a PMF with $\alpha=4$ and amplitude of
order $10^{-13}$ G at $1$ Mpc is constrained by CMB fluctuations
\citep{cmb-21}. If we constrain the PMF by the CMB polarization
plane rotation angle, \cite{WMAP}, then the limits on $B^{(\rm
eff)}$ range over $10^{-8} - 10^{-7}$ G for a PMF with spectral
index $n_B \in (-3,2)$, \cite{kah10}. Of course, the smoothed
amplitude $B_\lambda$ at 1 Mpc is very different if we compare an
observationally allowed scale-invariant spectrum with an
observationally-allowed $n_B=2$ spectrum. This has a very simple
explanation: to get $B^{\rm eff} \sim 10^{-9}$ G for $\alpha>-1$ a
lower value of $B_\lambda$ is required for $\lambda >\lambda_D$.

\subsection{Relic gravitational waves}

Another potentially interesting consequence of a PMF is relic
gravitational wave generation \citep{g1,gw,kcgmr08}. The amplitude
of these GWs is determined by the amount of MF energy density
present at the PT, and it is not affected by further MF damping.
There have been a number of studies of turbulence-generated GWs and
their detection prospects
\citep[see][]{kmk02,gkk07,kgr08,kks09,cds09, caprini09b}. The
spectrum of GWs is very sensitive to the temporal and spatial
characteristics of the source. Based on Figs.\ 3 and 4 we argue here
that the generation process is more effective within the PT
timescale. This is because the amplitude of the source drops
significantly outside the inertial range. The temporal decorrelation
of the turbulence results in faster decorrelation of smaller eddies.
Hence the main contribution to the GW signal comes from the largest
turbulent eddies, which can be associated with the length scale
$\lambda_0$. To justify this conclusion we re-derive here the
amplitude of direct-cascade MHD-turbulence-generated GWs. We do not
consider the magnetic and kinetic turbulence sources separately, as
previous work did \citep{kcgmr08}, but instead consider them
together and make use of the direct analogy between hydrodynamic and
magnetic turbulence when equipartition is reached, i.e. $v_0 \simeq
v_A$, where $v_0$ is the r.m.s.\ of the turbulence motion velocity.
Note that to establish equipartition a few turn-over times are
required \citep{kbtr10}. Hence, the stationary turbulence
approximation is increasingly valid as compared to previous
assumptions \citep{kmk02}. In contrast to earlier analysis
\citep{gkk07,kks09}, we derive\footnote{We make use of the
aeroacoustic approximation analogy described in detail in
\citet{gkk07}. The main assumption of our derivation is the long
enough duration of turbulence. To make the result more physical we
express all quantities in terms of the phase transition parameters.
} the GW amplitude using dimensionless quantities independent of the
expansion of the Universe, and directly relate it to the PT $\gamma$
parameter,
\begin{equation}
h_C(f) \simeq 1.2 \times 10^{-15} v_A \gamma^{5/2}
    \left(\frac{100\,{\rm GeV}}{T_*} \right)
    \left(\frac{100}{g_*}\right)^{1/3}\!\!
    \left(\frac{f}{f_H}\right)^{1/2}\!\! S^{1/2}(f), \label{Hc}
\end{equation}
where $f$ is the frequency now and $f_H =
\lambda_H^{-1} \simeq 1.6 \times 10^{-5}\,{\rm Hz}\,
({g_*}/{100})^{1/6} ({T_*}/{100\,{\rm GeV}})$  is the Hubble
frequency now. Here $S(f)$ is determined by MF statistical
properties,
\begin{equation}
S(f) = C_M^2 \int_{1}^{{\rm Re}^{3/4}}\!\!\frac{dx}{x^6}
    \exp\!\left(-\frac{f^2 \gamma^2}{f_H^2 v_A^2 x^{4/3}}\right)
    {\rm erfc} \!\left( -\frac{f \gamma}{f_H v_A x^{2/3}} \right),
    \label{H}
\end{equation}
where ${\rm erfc}(x)$ is the complementary error function defined as
$\mbox{erfc}(x) = 1 - \mbox{erf}(x)$, where $\mbox{erf}(x) =
\int_0^x dy \exp(-y^2)$ is the error function.  As expected, the
integral in Eq.~(\ref{H}) is dominated by the large scale ($x \simeq
1$) contribution. Equations (\ref{Hc})--(\ref{H}), presented here
for the first time in such a simply-interpreble form, allow us to
straightforwardly estimate the strength of the GW signal if the
energy density of the PMF is known. Also, Eq.~(\ref{H}) clearly
shows that the GW signal amplitude sensitively depends on $v_A$
which is one of main parameters when considering GW detection
prospects. This analytical estimate is in excellent agreement with
the numerical estimates of \citet{gkk07}.

The amplitude and the energy density of the GW are related through
\begin{equation} h_C(f) = 1.26 \times 10^{-18} \left( \frac{\rm
Hz}{f} \right) \left[ h_0^2 \, \Omega_{\rm GW}(f) \right]^{1/2},
\label{gw-amplitude}
\end{equation}
where $\Omega_{\rm GW}(f)$ is the GW spectral energy density
parameter. Using Eqs. (\ref{Hc}) and (\ref{gw-amplitude}), we have
\begin{eqnarray}
\Omega_{\rm GW}(f) h_0^2 =  2.3\times 10^{-4} v_A^2 {\gamma}^5
\left(\frac{100}{g_\star}\right)^{1/3}\!\!
\left(\frac{f}{f_H}\right)^3 \!\! S(f). \label{omega-g}
\end{eqnarray}
Integrating $\Omega_{\rm GW}(f)$ over frequency, it can be seen that
the efficiency of GW production is low, $\propto v_A^3 \gamma^2$,
with peak GW frequency for the EWPT being $f_{\rm peak} \simeq
\gamma^{-1} v_A \lambda_H^{-1}$ \citep{gkk07} and an additional peak
at $\lambda_H^{-1}$ for helical MHD turbulence \citep{kgr08}, but
the signal is potentially observable by LISA \citep{kcgmr08}. For
low frequencies $f \ll f_H$, $\Omega_{\rm GW} \propto f^3$ (also see
Caprini et al. 2009a,b), and for high frequencies $f\gg f_H$ there
is exponential damping. The peak amplitude $\Omega_{\rm GW} (f_{\rm
peak}) = 2.3 \times 10^{-4} v_A^5 \gamma^2 (g_\star/100)^{-1/3}$ and
is independent of $T_\star$. The analysis above shows that the main
contribution to the GW background comes from the EWPT and we can
ignore the expansion of the Universe when studying GW generation
(even for a helical PMF).

\section{Conclusion}

We have constrained a causally-generated PMF produced prior to BBN,
by using the BBN bound on the relativistic energy density during
nucleosynthesis. The above analysis is independent of what kind of
perturbations might be induced by the PMF, and whether the PMF acts
as a long or short duration source for those perturbations \citep[in
particular, no overproduction of GWs occurs, which could lead to a
strong constraint on the PMF amplitude,][]{gw,Caprini:2005ed}. We
can also constrain a PMF generated after nucleosynthesis, but still
during the radiation-dominated epoch, by requiring that the PMF
energy density not be the dominant component. Figure 1 shows the PMF
limits without accounting for damping or decay of the PMF. This is
because PMF large-scale decay is very model dependent, and
comprehensive analysis requires correctly accounting for the initial
conditions.

We have also argued that the direct use of a smoothed PMF
$B_\lambda$ may cause confusion. We instead propose using the PMF
energy density when deriving constraints from PMF cosmological
signatures. In particular, using $\xi_m (t) \propto t^{n_\xi}$
(where the correlation length is associated with the largest size MF
eddy length $\lambda_0 =2\pi/k_0$) together with ${\mathcal E}_M
\propto t^{-n_E}$, in the framework of ${\mathcal E}_M^{\rm LS}
\propto B_\lambda^2 (\lambda k_0)^{\alpha +1}$, implies that
$B_\lambda (t) \propto t^{((\alpha +1) n_\xi - n_E)/2}$, leading to
$B_\lambda$ increasing if the smoothing scale $\lambda $ is fixed,
while the real MF energy density is decreasing. It is obvious that
$B_\lambda$ increasing in time is not a physical effect. Another
advantage of using the PMF energy density is that the BBN limit does
not depend on the energy scale at generation ($T_\star$), even
though all characteristic length scales are strongly $T_\star$
dependent.

Free turbulence decays on large scales so, even without knowing the
exact scaling law, this allows us to conclude that after the PT ends
turbulence has been largely damped and so cannot produce
gravitational radiation at the same level as that from the PT
itself.  At length scales larger than $\lambda_0$, even when the PMF
energy is decaying slower than on small scales, the magnitude of the
source is much lower due to the spectral energy density scaling
($\propto k^{\alpha}$) and no significant increase of GW generation
efficiency can occur \citep[even for a long-duration source,][]
{caprini09b}. Accounting for this, the direct detection of relic GWs
will allow us to study the PT MHD turbulence picture, if a
sufficient fraction (1-10$\%$) of the thermal energy during the PT
is present in the form of MF energy density.

\acknowledgments We thank the anonymous referee for very useful
comments. We appreciate helpful comments from A.\ Brandenburg and
K.\ Jedamzik and useful discussions with C.\ Caprini, R.\ Durrer,
S.\ Huber, L.\ Kisslinger, A.\ Kosowsky, T.\ Stev\-ens, K.\
Subramanian, and T.\ Vachaspati. We acknowledge partial support from
GNSF grant ST08/4-422, and DOE grant DE-FG03-99EP41043 and NASA
grant NNXlOAC85G. T.K.\ acknowledges the ICTP associate membership
program, and NORDITA for hospitality during the Electroweak Phase
Transitions workshop.

\clearpage

\begin{figure}
\begin{center}
\includegraphics[width=6.0truein]{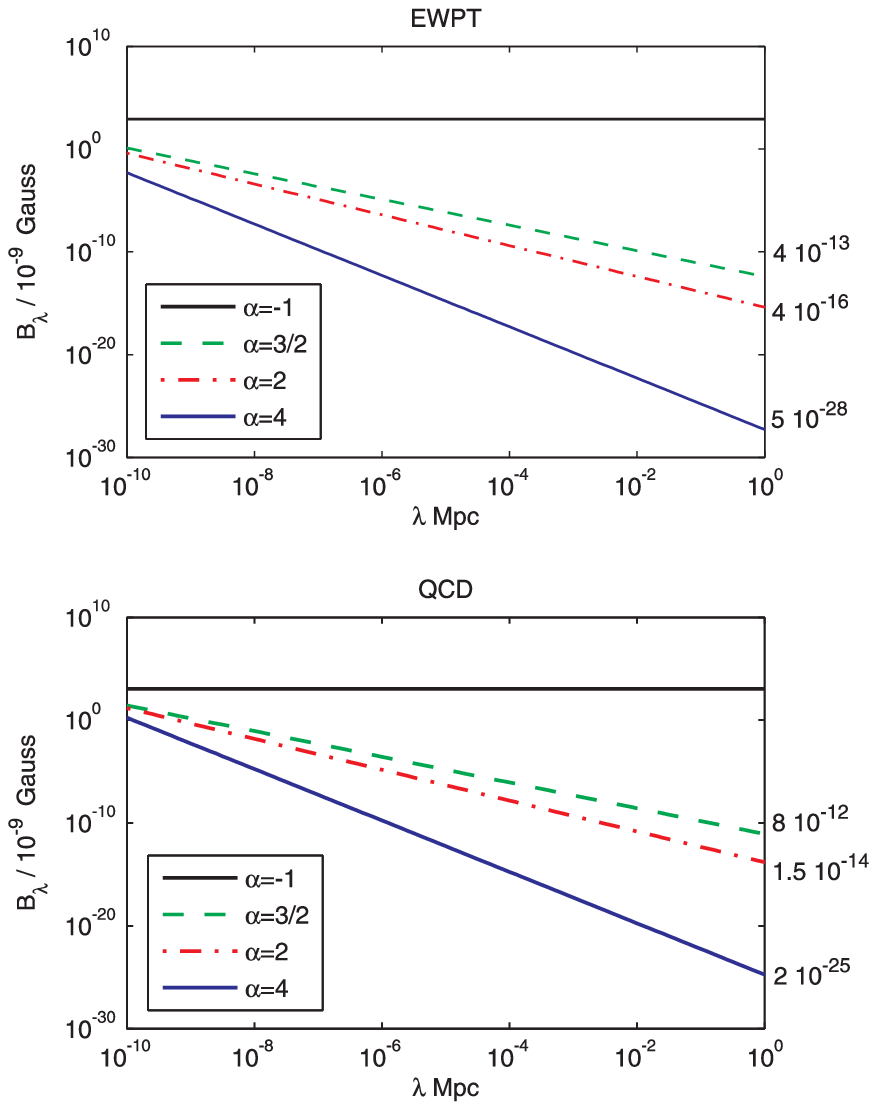}
\end{center}
\caption{The maximal allowed $B_\lambda$ for a PMF generated
during the EWPT (upper panel) with $T_\star=100$ Gev,
$g_\star=100$, and $\gamma=0.01$, and during the QCD PT (lower
panel) with $T_\star=0.15$ Gev, $g_\star=15$, and $\gamma=0.1$.
Limits display the effect of the expanding Universe. Realistic
nucleosynthesis bounds need to include effects of MF damping. }
\label{BL}
\end{figure}

\begin{figure}
\begin{center}
\includegraphics[width=6.0truein]{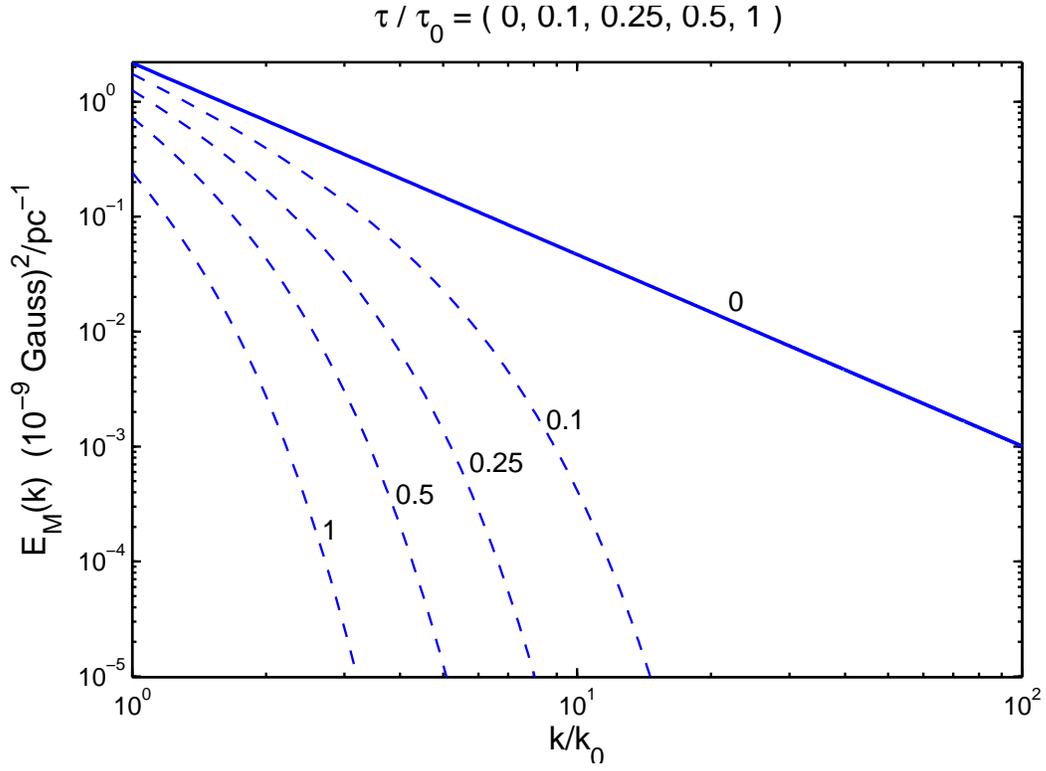}
\end{center}
\caption{Spectral energy density $E_M(\bar k)$ versus normalized
wavenumber $\bar k$ is shown during the freely decaying turbulence
decorrelation process at different points in time: $\tau/\tau_0$ =
0, 0.1, 0.25, 0.5, and 1.0. Inertial range of the direct cascade
turbulence spectrum is shown. } \label{Em_decorr}
\end{figure}

\begin{figure}
\begin{center}
\includegraphics[width=\columnwidth]{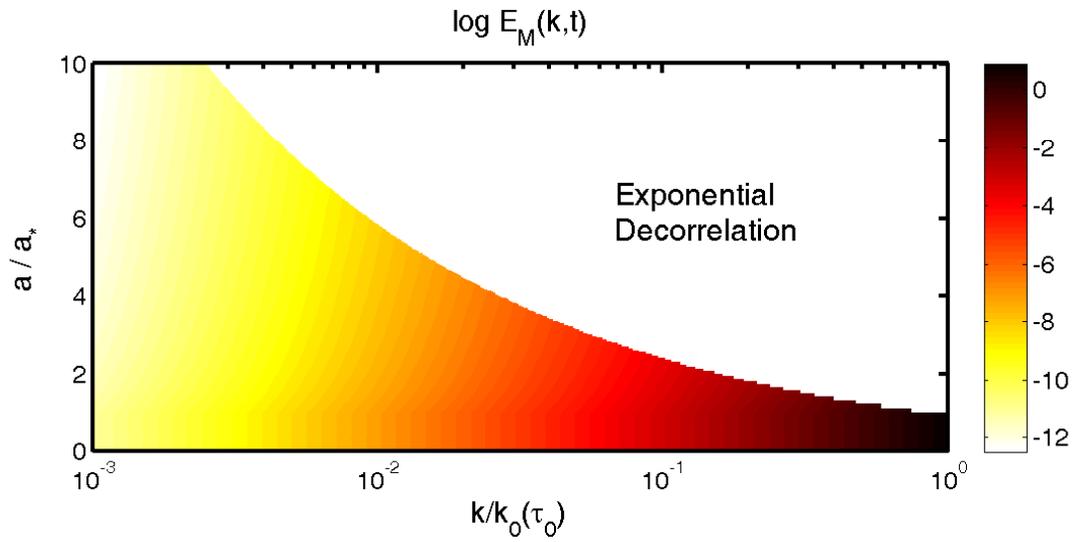}
\end{center}
\caption{Spectral energy surface of the large-scale non-helical
cascade turbulence ${\rm log} E_M(k,\eta)$. Free decay of turbulence
starts at $\tau/\tau_0=1$. } \label{Em_kt}
\end{figure}

\begin{figure}
\begin{center}
\includegraphics[width=\columnwidth]{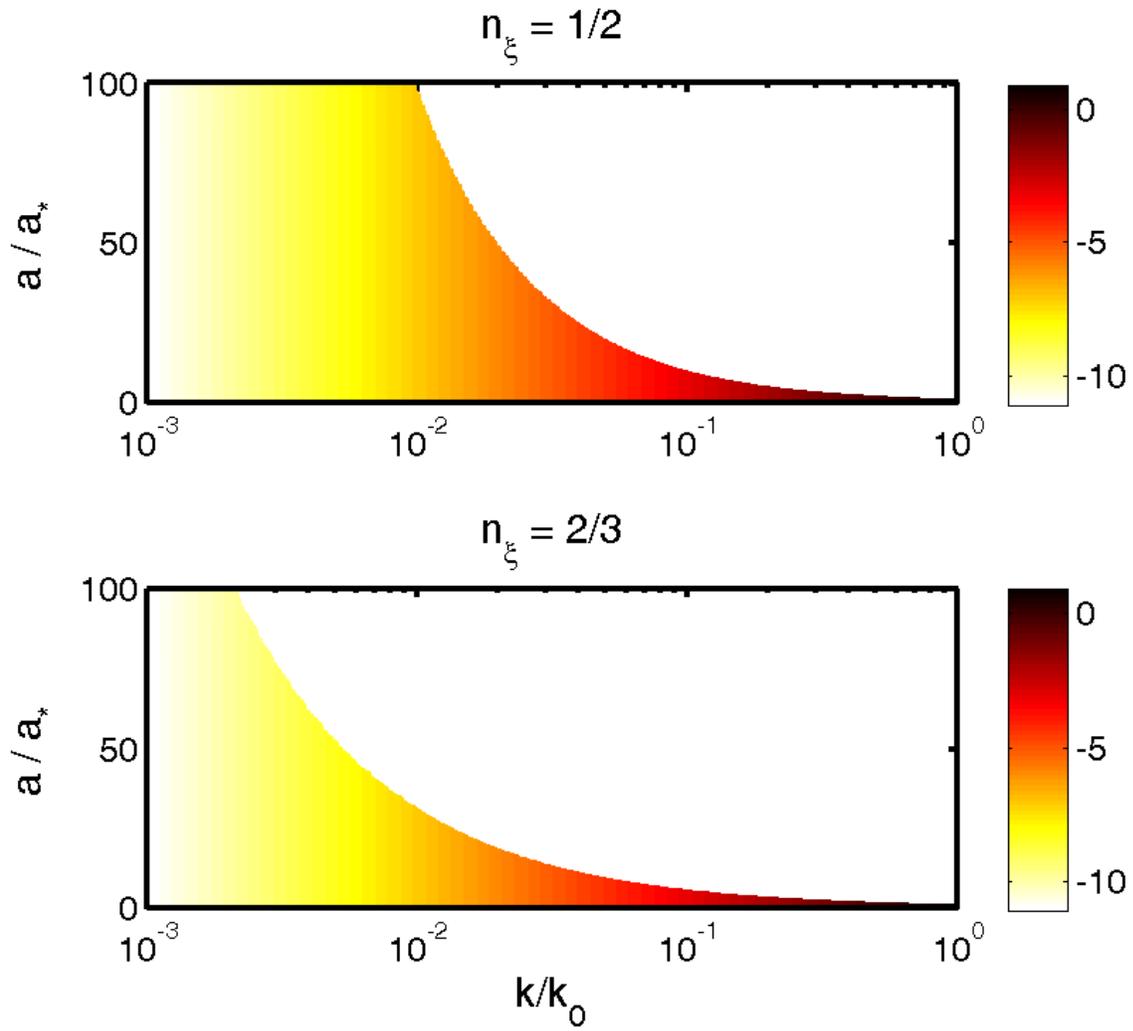}
\end{center}
\caption{Spectral energy surfaces of the large-scale helical
turbulence ${\rm log} E_M(k,\eta)$. The upper and lower panels
correspond to the $n_\xi=1/2$ and $n_\xi=2/3$ decay laws. }
\label{Em_aj}
\end{figure}

\end{document}